\documentclass{article}

\usepackage{arxiv}
\usepackage{mathptmx}
\usepackage{graphicx}
\usepackage{hyperref}      
\usepackage{amsmath,amsfonts,amssymb}
\usepackage{color}
\usepackage{braket}
\usepackage{float}
\usepackage{subfig}
\usepackage{cite}

\hypersetup{
   colorlinks=true,
   linkcolor=red,
  filecolor=cyan,      
    urlcolor=black,
   citecolor=blue}

\title{ Engineering two-mode squeezed vacuum-like states by using the Bohm potential}
\author{ Héctor M. Moya-Cessa\\Instituto Nacional de Astrofísica Óptica y Electrónica\\Calle Luis Enrique Erro No. 1, Santa María Tonantzintla, Pue., 72840, Mexico.\\
\And Felipe A. Asenjo
\\ Facultad de Ingeniería y Ciencias,
Universidad Adolfo Ibáñez, Santiago 7491169, Chile.
  \\ \And Sergio A. Hojman\\
  Departamento de Ciencias, Facultad de Artes Liberales,
Universidad Adolfo Ibáñez, Santiago 7491169, Chile.\\
Departamento de Física, Facultad de Ciencias, Universidad de Chile,
Santiago 7800003, Chile.\\
Centro de Recursos Educativos Avanzados,
CREA, Santiago 7500018, Chile.
\\  \And Francisco Soto-Eguibar \\
 Instituto Nacional de Astrofísica Óptica y Electrónica\\Calle Luis Enrique Erro No. 1, Santa María Tonantzintla, Pue., 72840, Mexico.\\}

\begin{document}

\maketitle

\begin{abstract}
We show that two-mode squeezed vacuum-like states may be engineered in the Bohm-Madelung formalism by adequately choosing the phase of the wavefunction. The difference between our wavefunction and the one of the squeezed vacuum states is given precisely by the phase we choose.
\end{abstract}


\section{Introduction}
Engineering non-classical states of a quantum mechanical system has been a main goal since the discovery of quantum mechanics \cite{Schroedinger}. Non-classical states are of interest as they show less fluctuations in a given canonical variable (at the expense of the other) than that of coherent states that have fluctuations referred to as the standard quantum limit.\\
In coupled harmonic oscillators, one of the most studied non-classical states that most labs try to generate are the so-called two-mode squeezed vacuum states \cite{TMSV}  because of their several applications; among them, the enhancement of two-photon spectroscopy \cite{Prajapati}. Generation of two mode squeezed vacuum states and in particular of the superposition of two-mode squeezed states \cite{Yuen,Caves,Loudon} may be used for quantum information processing and quantum sensing \cite{TMarxiv}. Coupled time dependent harmonic oscillators may be studied by using Ermakov-Lewis \cite{Lewis,Guasti,Pedrosa,Ray} (invariant) techniques as done by several authors \cite{Guedes,Recamier,Iran}.\\
In this manuscript we look at the problem of generation of two-moded squeezed states by shaping the phase of the wavefunction. In fact, methods to tailor the wavefunction of a quantum state based on quantum teleportation have been studied \cite{Asav}. In particular, we will use  an operator approach to the Madelung-Bohm theory that, to the best of our knowledge, have not been approached using either operator techniques or group theory. We solve the Madelung-Bohm continuity or probability conservation equation via such approach and, by using a particular form of the wavefunction's phase, i.e., a tailored wavefunction, we are able to show that we may generate non-classical states of two-coupled time dependent harmonic oscillators, namely, two-mode squeezed vacuum states.

\section{Madelung-Bohm equations}
Let us consider the Schrödinger equation
\begin{equation}\label{SE}
    i\frac{\partial \psi(x,y,t)}{\partial t}= -\frac{1}{2m}\frac{\partial^2 \psi(x,y,t)}{\partial x^2}-\frac{1}{2m}\frac{\partial^2 \psi(x,y,t)}{\partial y^2}+V(x,y,t)\psi(x,y,t)
\end{equation}
with $\psi(x,y,t)$ the wavefunction of the quantum mechanical system, $V(x,y,t)$ an external arbitrary potential which can be time dependent and where we set $\hslash=1$. We adopt the Madelung-Bohm approach to quantum mechanics \cite{madel,bohm,durr} and write the wavefunction in terms of a polar decomposition 
\begin{equation}\label{psi}
\psi(x,y,t)=A(x,y,t)e^{i S(x,y,t)}
\end{equation}
which gives rise to two equations, for the real and imaginary components of the Schrödinger equation, namely, a modified Hamilton--Jacobi equation
\begin{equation} \label{H-J}
\frac{1}{2m}S_x^2+\frac{1}{2m}S_y^2+V_B+V+S_t=0,
\end{equation}
and the continuity (probability conservation) equation
\begin{equation}\label{Prob}
\frac{\partial A}{\partial t}+\frac{1}{m} 
\left( S_x\frac{\partial A}{\partial x}
+S_y \frac{\partial A}{\partial y}\right)
+\frac{1}{2m} \left(S_{xx} A + S_{yy} A\right)=0,
\end{equation}
where the sub-indices represent partial derivatives, where we have omitted the dependence in $x,\;y,\;t$ of all functions, and where we have introduced the Bohm potential, which is defined as 
\begin{equation}\label{bohmpot}
V_B(x,y,t)=-\frac{1}{2m A(x,y,t)}\left[ \frac{\partial^2 A(x,y,t)}{\partial x^2}
+\frac{\partial^2 A(x,y,t)}{\partial y^2}\right].
\end{equation}

\section{Operator solution to Bohm potential in 3D}
The probability equation \eqref{Prob} may be rewritten as a Schrödinger-like equation
\begin{equation}\label{Operator}
\frac{\partial A}{\partial t}=-\frac{1}{2m} \left(2 S_x \frac{\partial }{\partial x}+2 S_y \frac{\partial }{\partial y}+S_{xx} +S_{yy} \right) A.
\end{equation}
To obtain the two-mode squeezed vacuum-like states, we set the phase
\begin{equation}\label{fase}
    S(x,y,t)=m{\dot{\nu}}(t)[r\frac{x^2+y^2}{2}+xy]+\mu(t),
\end{equation} 
where $r$ is a parameter that will be determined later. Thus, we rewrite Eq.~\eqref{Operator} as
\begin{equation}\label{OperatorI}
\partial_t A=-{i}\dot{\nu}\left[(rx+y)\hat{p}_x+(ry+x)\hat{p}_y]-ir \right]A,
\end{equation}
by using the momentum operators $\hat{p}_x=-i\frac{\partial }{\partial x}$ and $\hat{p}_y=-i\frac{\partial }{\partial y}$. Its formal solution is
\begin{align}\label{opsum}
A(x,y,t)&=e^{-i\nu\left[(rx+y)\hat{p}_x+(ry+x)\hat{p}_y-ir \right]}A_0(x,y)
\nonumber \\ &
=e^{-i\nu\left[r\left(x\hat{p}_x+y\hat{p}_y\right)+\left(x\hat{p}_y+y\hat{p}_x \right)-ir\right] } A_0(x,y)
\end{align}
where $A_0(x,y)=A(x,y,t=0)$ is the initial condition which determines that we must have $\nu(t=0)=0$.\\
It may be seen from the above equation that the relevant operators are $y\hat{p}_x+x\hat{p}_y$ and $x\hat{p}_x+y\hat{p}_y$ (a product of squeeze operators \cite{Wheeler}) and as they commute, we may write \eqref{opsum} as
\begin{equation}\label{opsumI}
    A(x,y,t)=e^{-r\nu}e^{-{i}r{\nu}(x\hat{p}_x+y\hat{p}_y)}e^{-{i}{\nu}(y\hat{p}_x+x\hat{p}_y)}A_0(x,y).
\end{equation}
We introduce now the annihilation and creation operators for both $x$ and $y$ dimensions as usual
\begin{equation}
   \hat{a}=\frac{x+i\hat{p}_x}{\sqrt{2}}, \qquad   \hat{a}^{\dagger}=\frac{x-i\hat{p}_x}{\sqrt{2}},
\end{equation}
and
\begin{equation}
   \hat{b}=\frac{y+i\hat{p}_y}{\sqrt{2}}, \qquad   \hat{b}^{\dagger}=\frac{y-i\hat{p}_y}{\sqrt{2}}.
\end{equation}
and we can cast the formal solution Eq. \eqref{opsumI} as
\begin{align}\label{opsumII}
A(x,y,t)=&\exp\left(-2r\nu\right) 
\exp\left[r\frac{\nu}{2}\left(\hat{a}^2-\hat{a}^{\dagger 2}+\hat{b}^2-\hat{b}^{\dagger 2} \right)  \right]
\exp\left[\nu\left(\hat{a}^\dagger\hat{b}^\dagger-\hat{a}\hat{b} \right)  \right] A_0(x,y);
\end{align}
it is worth to note two aspects: first, that the operator $\exp\left[r\frac{\nu}{2}\left(\hat{a}^2-\hat{a}^{\dagger 2}+\hat{b}^2-\hat{b}^{\dagger 2} \right)  \right]$ is the well known squeeze operator in two dimensions, and second, that the first acting operator, the one most to the right, does not depend on $r$. Thus, if we understand the action of this last operator, we have a solution to our problem.\\
In Appendix \ref{apa}, we show how to factorize the evolution operator $e^{\nu(\hat{a}^{\dagger}\hat{b}^{\dagger}-\hat{a}\hat{b})}$, and using that factorization we get
\begin{align}\label{opsumIII}
A(x,y,t)=&\frac{\exp\left(-2r\nu\right) }{\cosh\nu}
\exp\left[r\frac{\nu}{2}\left(\hat{a}^2-\hat{a}^{\dagger 2}+\hat{b}^2-\hat{b}^{\dagger 2} \right)  \right]
\nonumber \\ & 
\exp\left[\tanh\left(\nu \right) \hat{a}^\dagger \hat{b}^\dagger\right] 
\exp\left[-\ln\left(\cosh \nu \right) \left( \hat{a}^\dagger\hat{a}+\hat{b}^\dagger\hat{b}\right)  \right] 
\exp\left[-\tanh\left(\nu \right) \hat{a}\hat{b} \right] A_0(x,y).
\end{align}
We would like to underline that since we have set the phase of the wave function, Eq.~\eqref{fase}, and we have also found the amplitude of the same wave function, Eq.~\eqref{opsumIII}, we can get the external potential $V(x,y,t)$ clearing it from Eq.~\eqref{H-J}; below we will get an explicit expression for a given initial condition. We want also to remark that the wave function \eqref{psi}, with Eq.~\eqref{fase} and Eq.~\eqref{H-J}, is a solution of the Schr\"odinger equation \eqref{SE} with the corresponding external potential explained in the previous paragraph. 

\section{Two-mode squeezed vacuum states}
Let us consider as initial condition the separable initial state $A_0(x,y)=\varphi_0(x)\varphi_0(y)$ with
\begin{equation}
\varphi_n(\eta)=\frac{1}{\sqrt{2^n n!\sqrt{\pi}}}e^{-\frac{\eta^2}{2}}H_n(\eta),
\end{equation}
being $H_n(\eta)$ the Hermite polynomial of order $n$ and which are neither more nor less than the proper functions of the harmonic oscillator Hamiltonian.\\

\subsection{Case $r=0$}
As we explain above, if we establish $A(x,y,t)$ for $r=0$, we have the complete solution as soon as the other operator is just the squeeze operator. Thus, if we make $r=0$ in \eqref{opsumIII} and we introduce there the initial condition $A_0(x,y)=\varphi_0(x)\varphi_0(y)$, the first and second acting operators will give simply $A_0(x,y)$ as $\hat{a}\varphi_0(x)=0$ and $\hat{b}\varphi_0(y)=0$. Therefore, if we use the fact that the operators $\hat{a}$ and $\hat{b}$ are the ladder operators for the functions $\varphi_n(x)$ and $\varphi_n(y)$, respectively, we may write
\begin{align} 
    A(x,y,t) = & \frac{1}{\cosh\nu}e^{\tanh(\nu) \hat{a}^{\dagger}\hat{b}^{\dagger}} \varphi_0(x)\varphi_0(y)
    \nonumber \\ 
    = & \frac{1}{\cosh\nu}\sum_{n=0}^{\infty}\tanh^n(\nu) \varphi_n(x)\varphi_n(y),
\end{align}
that is a highly entangled state for $\nu > 0$.\\
Using Mehler's formula~\cite{Mehler,erdelyi,stone,andrews}
\begin{equation}
    \sum_{n=0}^{\infty}\frac{1}{n!} \left(\frac{\rho}{2}\right)^n H_n(x)H_n(y)=\frac{1}{\sqrt{1-\rho^2}}
    \exp\left[-\frac{\rho^2(x^2+y^2)-2\rho xy}{1-\rho^2}\right],
\end{equation}
and taking $\rho=\tanh \nu$, we finally obtain
\begin{equation}
   A(x,y,t) =\frac{1}{\sqrt{\pi}}
   \exp\left[-\frac{x^2+y^2}{2}-\frac{(x^2+y^2)\tanh^2\nu-2xy\tanh \nu }{\text{sech}^2 \nu}\right].
\end{equation}

\subsection{Case $r\ne 0$}
In the case $r\ne 0$, we need to apply the squeeze operator given in (\ref{opsumI}) to the above equation, which readily gives
\begin{equation} \label{0190}
   A(x,y,t) =\frac{1}{\sqrt{\pi}} \exp\left(-r \nu\right)
   \exp\left[- \frac{x^2+y^2}{2}e^{-2r\nu}-\frac{(x^2+y^2)\tanh^2\nu-2xy\tanh \nu }{\text{sech}^2\nu}e^{-2r\nu}\right].
\end{equation}
From expression \eqref{bohmpot}, we can now write explicitly the Bohm potential 
\begin{align}\label{0200}
    V_B(x,y,t)=-\frac{1}{2m} \left[
    \left(x^2+y^2\right) \cosh (4 \nu ) e^{-4 \nu  r}
    -2 x y \sinh (4 \nu ) e^{-4 \nu  r}
    -2 \cosh (2 \nu ) e^{-2 \nu  r}
    \right],
\end{align}
and clearing $V(x,y,t)$ from \eqref{H-J} the external potential
\begin{align}\label{0210}
    V(x,y,t)=&\frac{1}{2} \left(x^2+y^2\right) \left[-m \left(r^2+1\right) \dot{\nu}^2-m   r \ddot{\nu}-2 \cosh (4 \nu ) e^{-4 \nu   r}\right]
    \nonumber \\ &
    +x y \left[-2 m r \dot{\nu}^2-m \ddot{\nu}+2 \sinh (4 \nu ) e^{-4 \nu  r}\right]+2 \cosh (2 \nu ) e^{-2 \nu  r}-\dot{\mu},
\end{align}
that correspond to a time dependent potential for two coupled time dependent harmonic oscillators \cite{Urzua}.

Thus, the wavefuntion \eqref{psi} with phase \eqref{fase} and amplitude \eqref{0190} is solution of the Schr\"odinger equation \eqref{SE} with external potential \eqref{0210}.\\
Both potentials, the Bohm \eqref{0200} and the external \eqref{0210}, are paraboloids in the space. The levels curves of both potentials are conics sections. Using elementary techniques of analytic geometry, we find that the discriminant of the level curves of the Bohm potential is $D=-\frac{e^{-10 r \nu (t)} \cosh \left[ 2 \nu (t)\right] }{4 m^3}$ which is always different from zero, so the level curves are always non-degenerated conics, and the minor of the element 33 of the matrix of the corresponding quadratic form is $\frac{e^{-8 r \nu (t)}}{4 m^2}$ which is always positive, and thus the level curves will be in all cases ellipses. So independently of $\nu$, $\mu$ and of the values of the parameters, the Bohm potential will have always the same qualitative behavior; it is also worth to note that we have the same conduct for $r\neq0$ than for $r=0$.\\
In the case of the external potential, Eq.~\eqref{0210}, the situation is far more complicated and the level curves can be any kind of conic included degenerated cases which can be parallel straight lines; this open the possibility of the appearing of some kind of phase transitions, but we will not explore that possibility in this work.

\subsection{A first example }
Let us consider $m=1,\;r=0,\;\nu(t)=t$ and $\mu(t)=0$. The wave function results to be
\begin{equation}
    \psi(x,y,t)=\frac{1}{\sqrt{\pi}}\exp \left[-\frac{1}{2} \cosh (2 t) \left(x^2+y^2\right)+x y \sinh (2 t)+i x y\right],
\end{equation}
\begin{figure}[H]
	\subfloat[$t=0$ \label{fig1a}]
	{\includegraphics[width=0.45\textwidth,height=0.45\textheight,keepaspectratio]{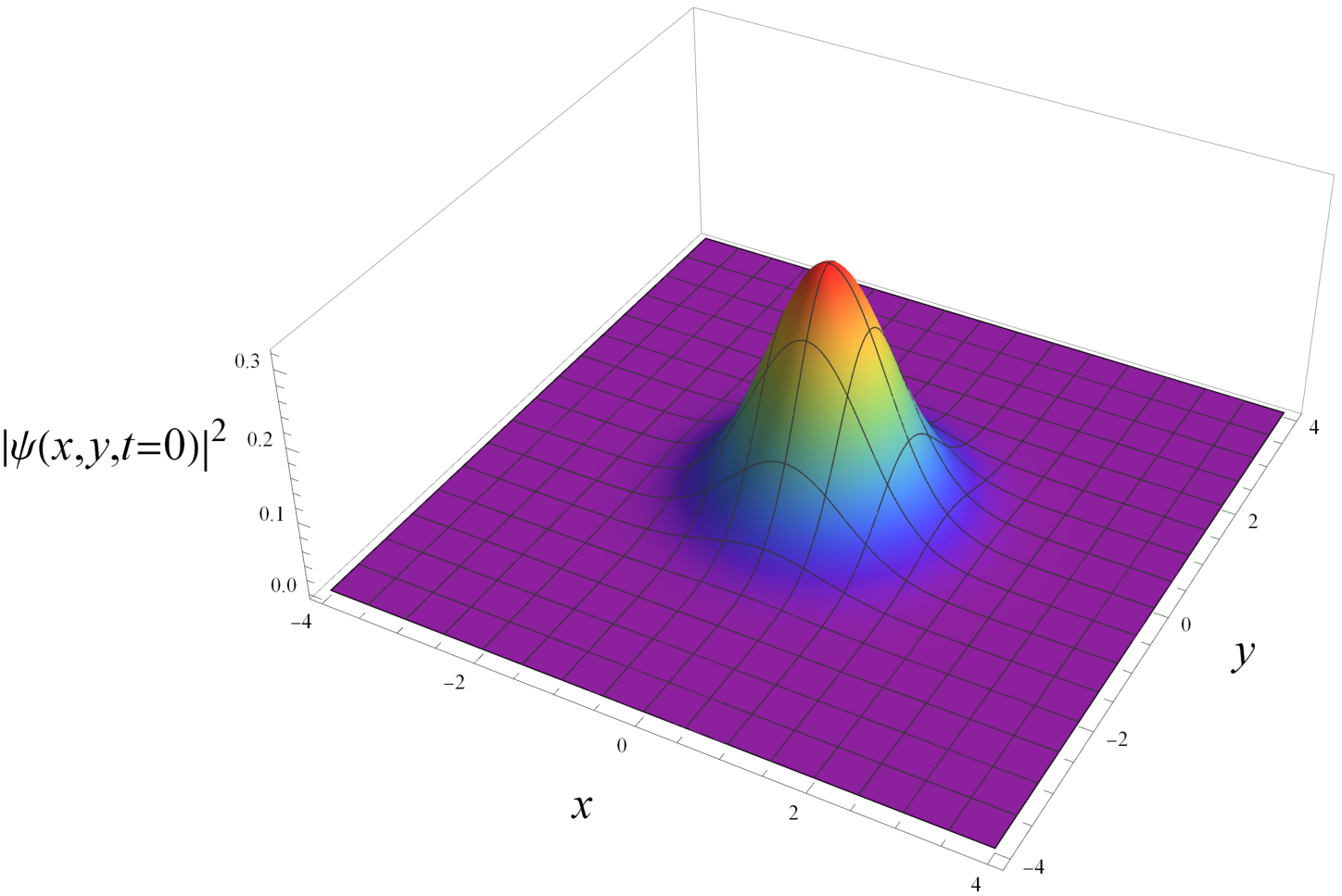}}
	\hfill
	\subfloat[$t=1$ \label{fig2a}]
	{\includegraphics[width=0.45\textwidth,height=0.45\textheight,keepaspectratio]{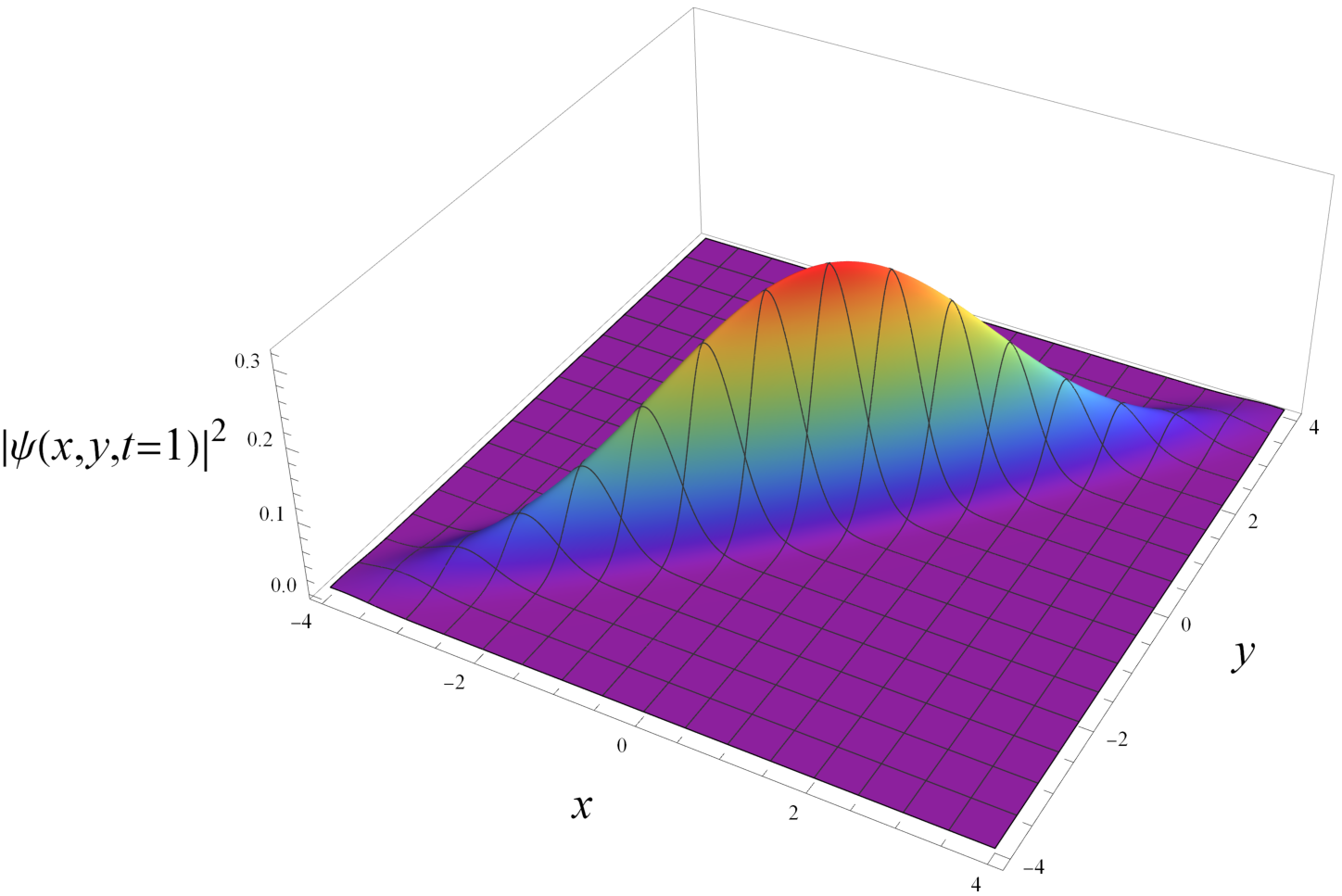}}
	\\
	\subfloat[$t=2$ \label{fig3a}]
	{\includegraphics[width=0.45\textwidth,height=0.45\textheight,keepaspectratio]{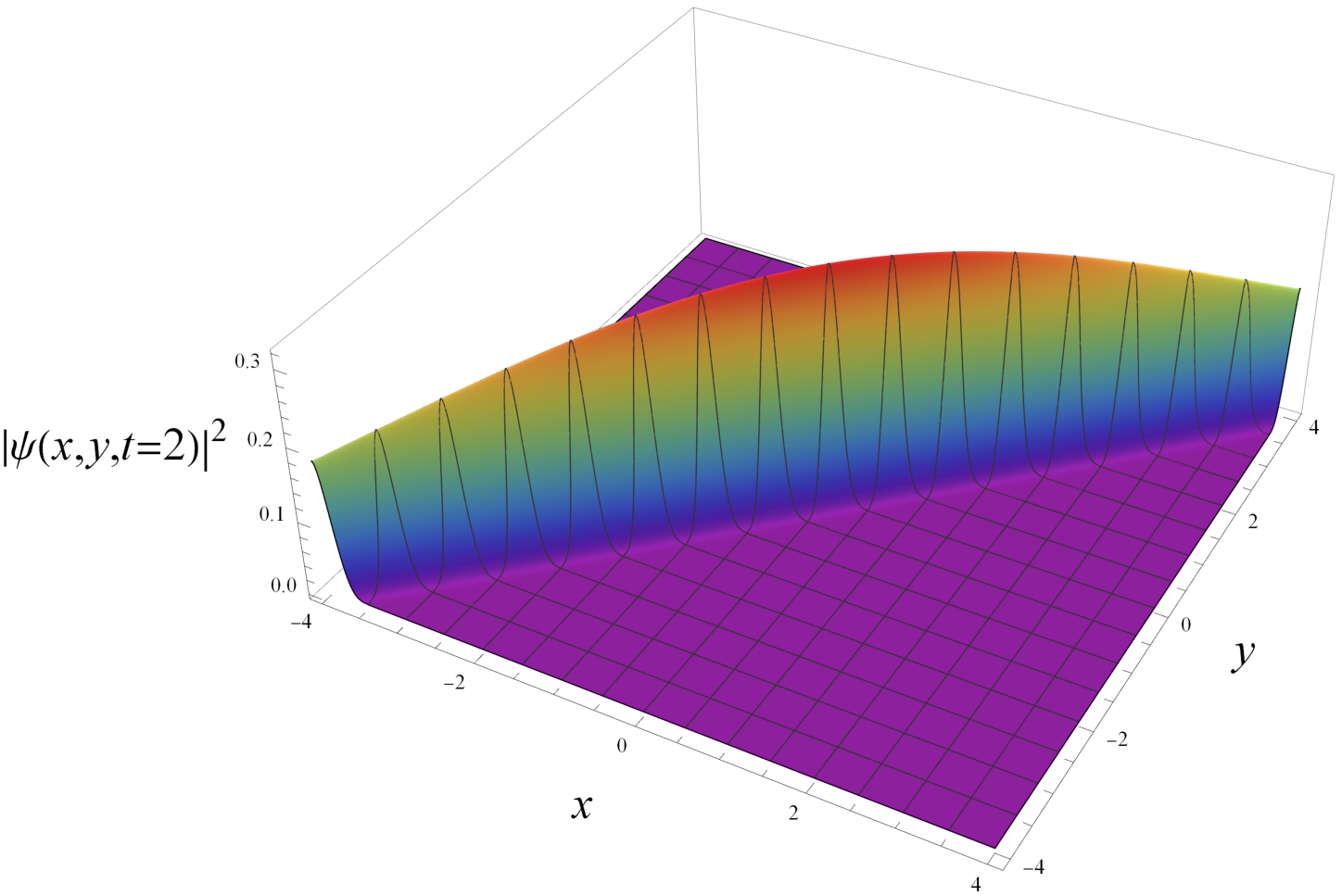}}
	\hfill
	\subfloat[$t=3$ \label{fig4a}]
	{\includegraphics[width=0.45\textwidth,height=0.45\textheight,keepaspectratio]{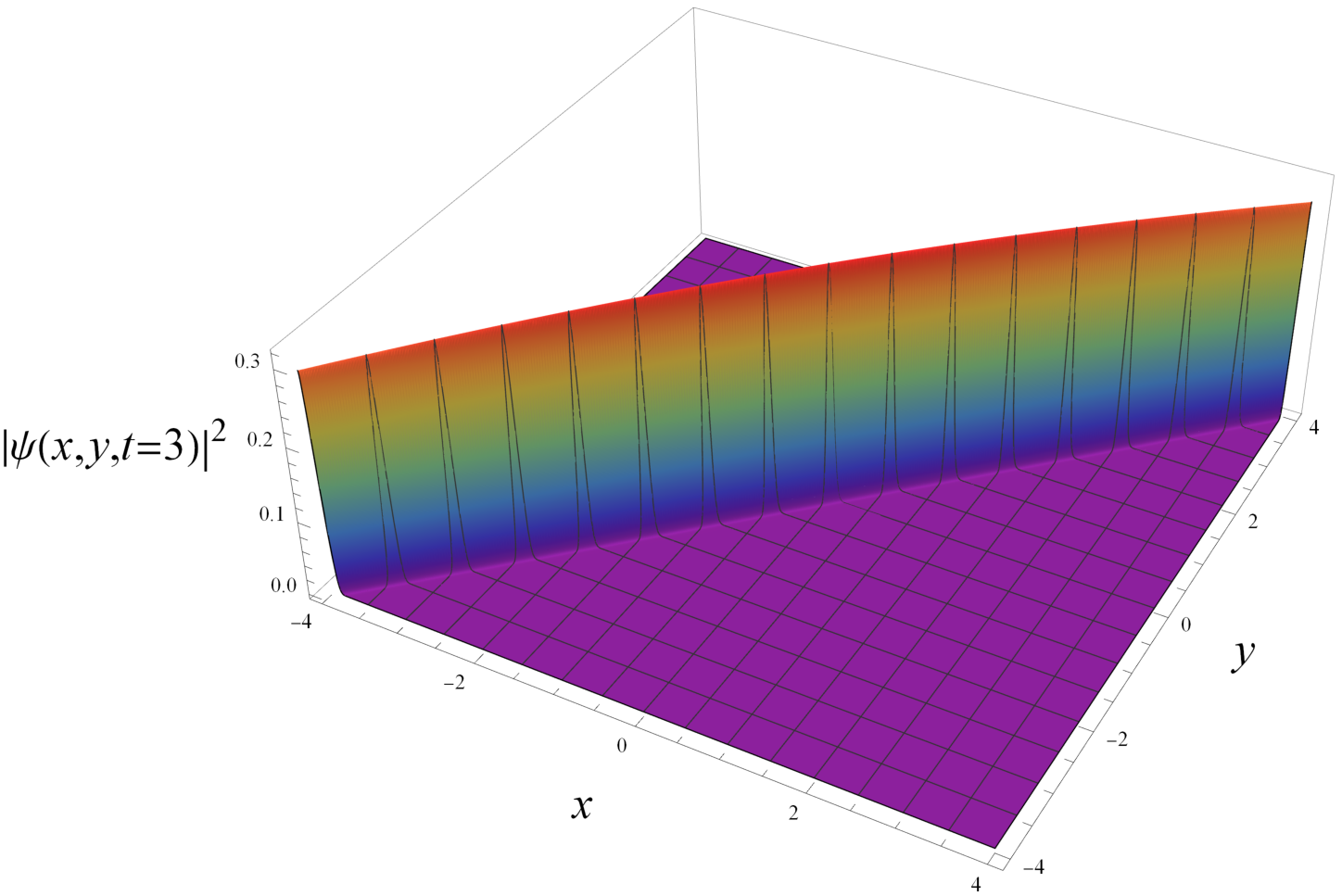}}
	\caption{The square of the absolute value of the wave function $|\psi(x,y,t)|^2$ for $m=1,\;r=0,\;\nu(t)=t,\;\mu(t)=0$ at different times.}
	\label{fig1}
\end{figure}
the Bohm potential is
\begin{equation}
    V_B(x,y,t)=-\frac{1}{2} (x^2+y^2) \cosh (4 t)+x y \sinh (4 t)+\cosh (2 t),
\end{equation}
and the external potential is
\begin{equation}
    V(x,y,t)=\frac{x^2+y^2}{2} \left[\cosh (4 t)-1\right]-x y \sinh (4 t)-\cosh (2 t).
\end{equation}
\textcolor{black}{In Fig.~\ref{fig1}, we plot the square of the wavefunction for different times. Starting from a (radial) Gaussian function with no preferred angle, it squeezes to give a well defined angle. Squeezing occurs for a variable at $45$ degrees where the wavefunction produces a well defined position}.

\subsection{A second example}
In this second example, we set $m=1,\;r=1,\;\nu(t)=t^2$ and $\mu(t)=0$, and we obtain for the wave function
\begin{equation}
    \psi(x,y,t)=\frac{1}{\sqrt{\pi}}
    \exp \left[-e^{-2 t^2} \cosh\left(2 t^2\right)\frac{x^2+y^2}{2} +e^{-2 t^2}\sinh \left(2 t^2\right) x y+ i t \left(x+y\right)^2 -t^2\right],
\end{equation}
\begin{figure}[H]
	\subfloat[$t=0$ \label{fig5a}]
	{\includegraphics[width=0.45\textwidth,height=0.45\textheight,keepaspectratio]{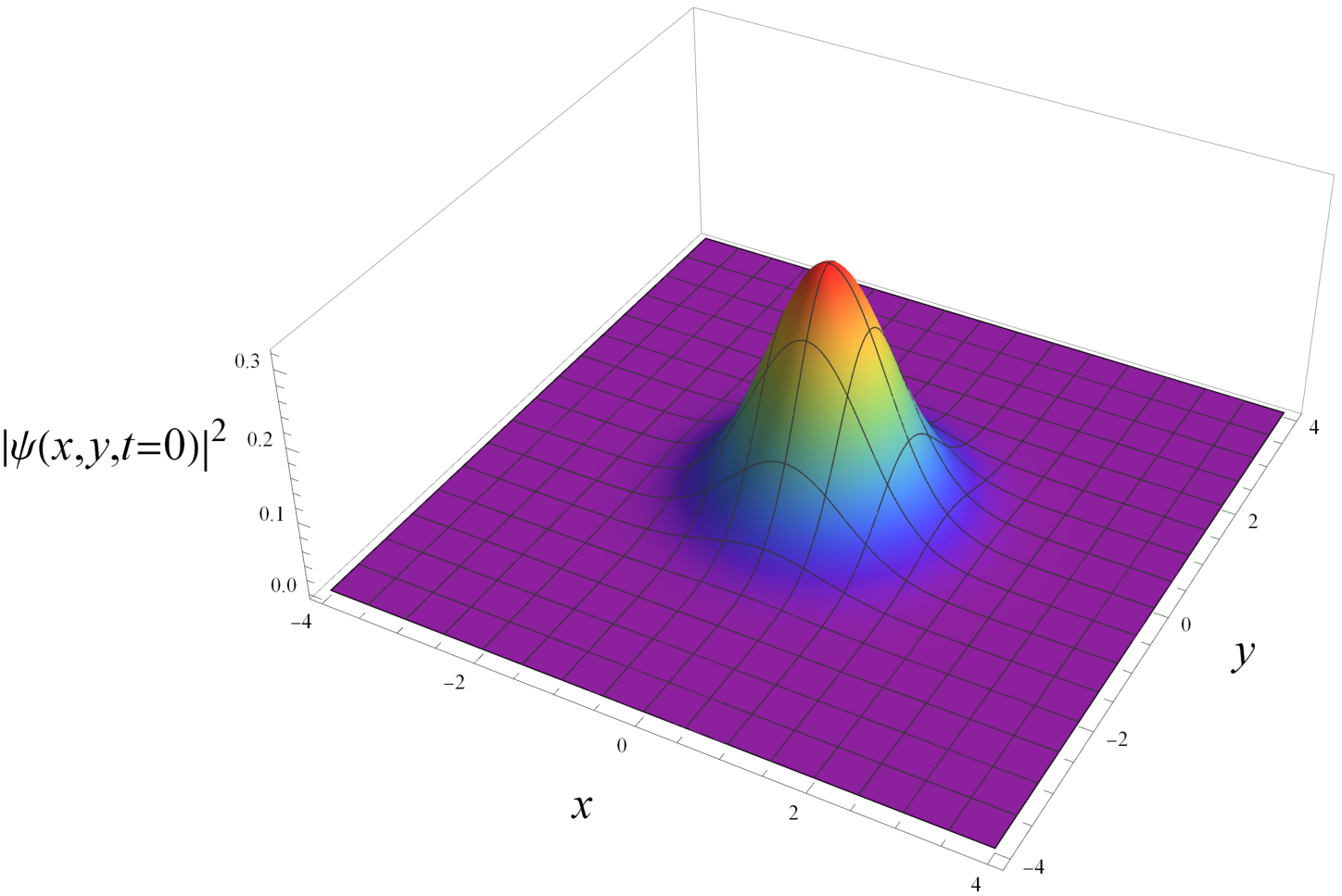}}
	\hfill
	\subfloat[$t=1$ \label{fig6a}]
	{\includegraphics[width=0.45\textwidth,height=0.45\textheight,keepaspectratio]{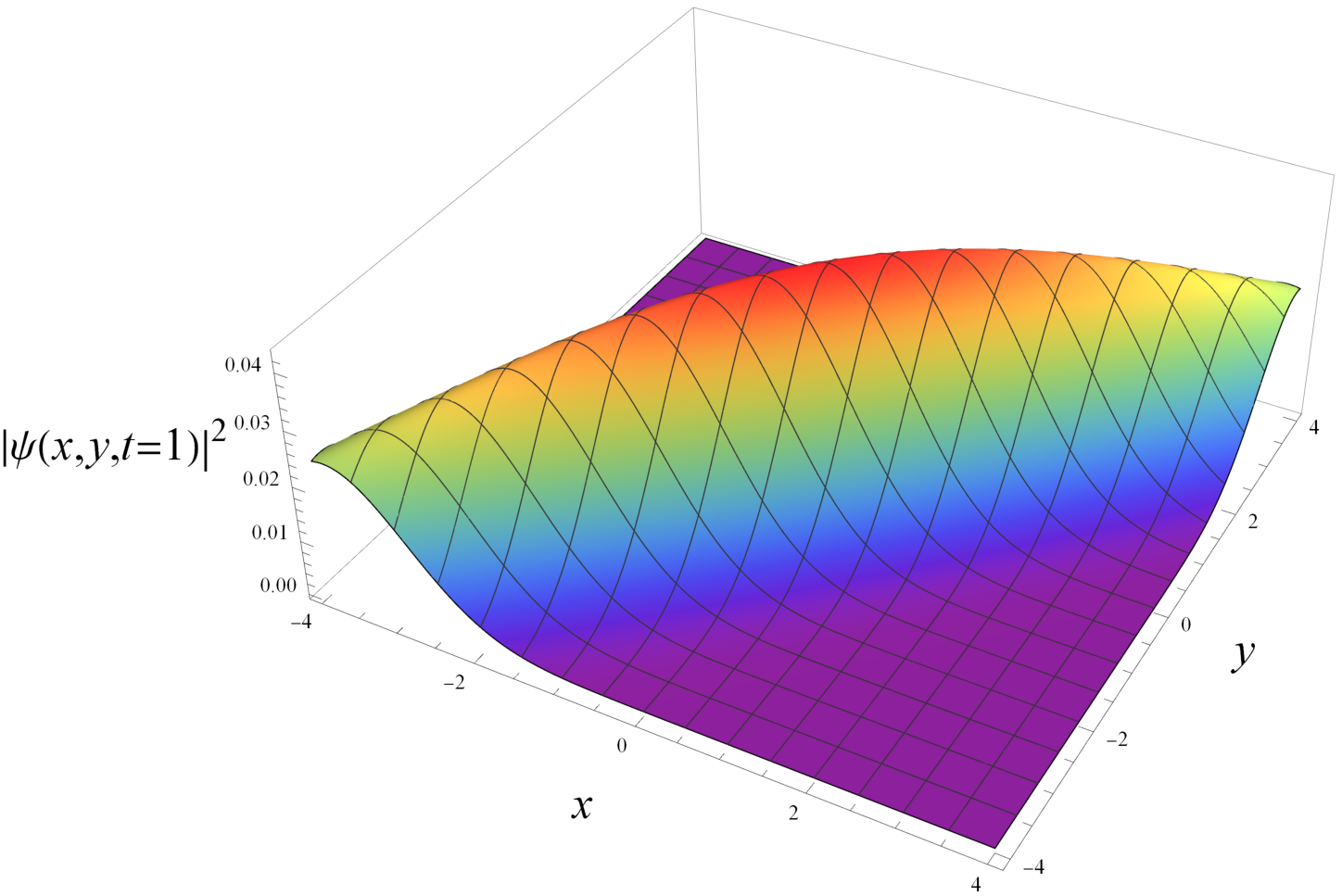}}
	\\
	\subfloat[$t=2$ \label{fig7a}]
	{\includegraphics[width=0.45\textwidth,height=0.45\textheight,keepaspectratio]{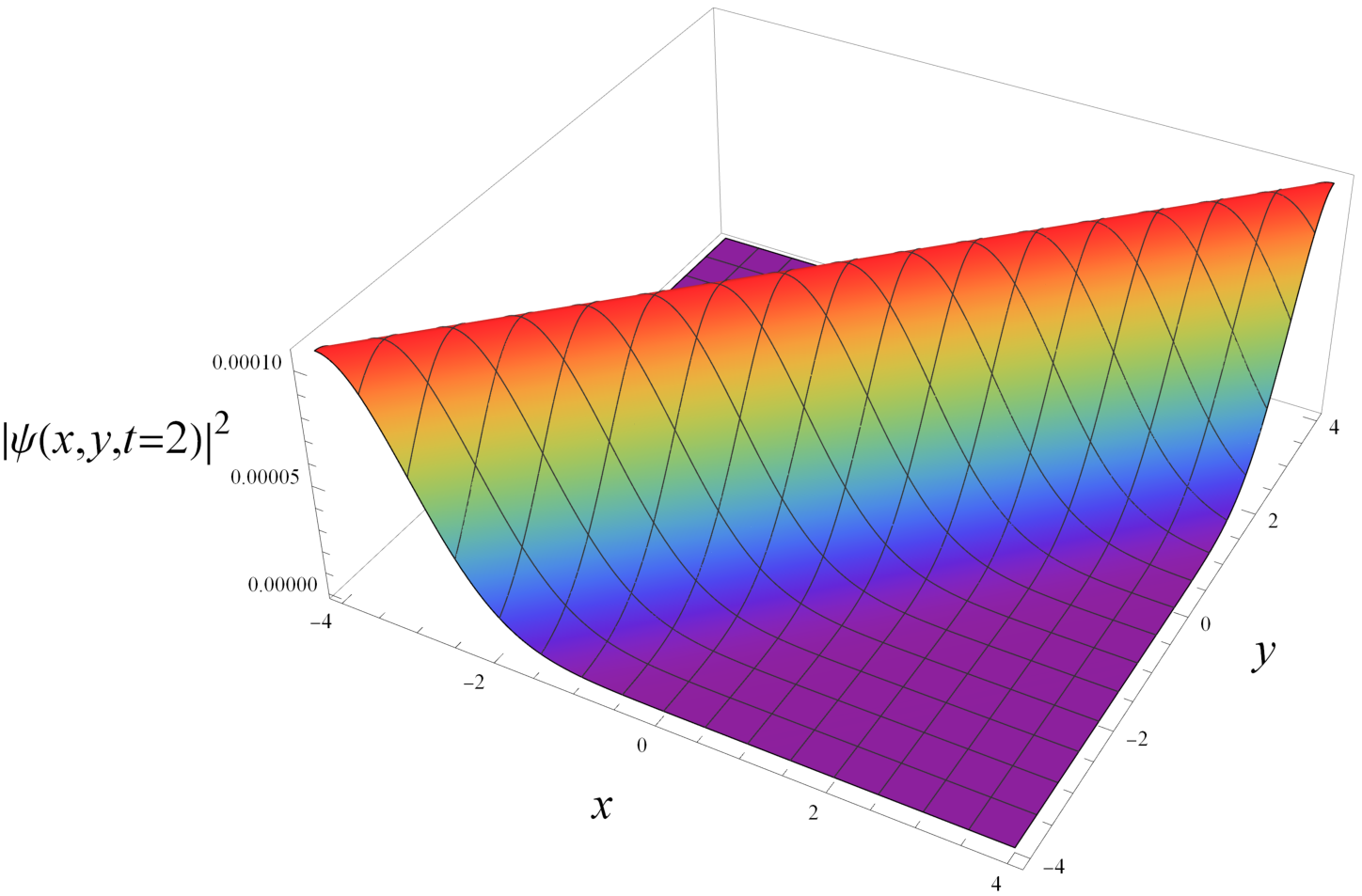}}
	\hfill
	\subfloat[$t=3$ \label{fig8a}]
	{\includegraphics[width=0.45\textwidth,height=0.45\textheight,keepaspectratio]{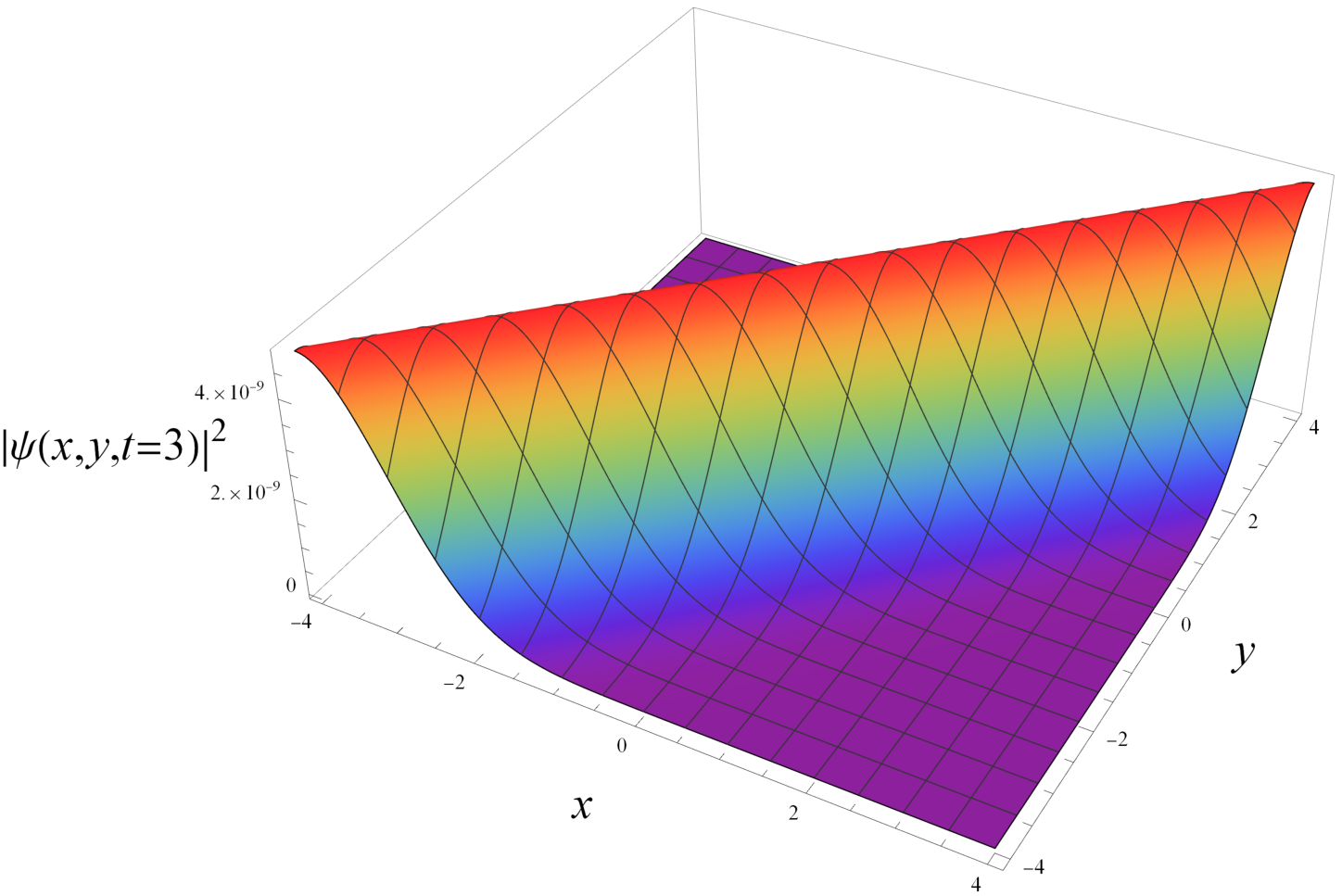}}
	\caption{The square of the absolute value of the wave function $|\psi(x,y,t)|^2$ for $m=1,\;r=1,\;\nu(t)=t^2,\;\mu(t)=0$ at different times.}
	\label{fig4}
\end{figure}
for the Bohm potential
\begin{equation}
    V_B(x,y,t)=-\frac{1}{2} e^{-4 t^2} (x^2+y^2) \cosh \left(4 t^2\right)+e^{-4 t^2} x y \sinh \left(4 t^2\right)+e^{-2 t^2} \cosh \left(2 t^2\right),
\end{equation}
and for the external potential
\begin{equation}
    V(x,y,t)=\left[-4 t^2+\frac{1}{2} e^{-4 t^2} \cosh \left(4 t^2\right)-1\right] \left(x^2+y^2\right)+ \left[-8 t^2-e^{-4 t^2} \sinh \left(4 t^2\right)-2\right]x y-e^{-2 t^2} \cosh \left(2 t^2\right).
\end{equation}
\textcolor{black}{In Fig.~\ref{fig4}, we plot the square of the wavefunction for different times. Starting from a (radial) Gaussian function with no preferred angle, it also  squeezes to give a well defined angle. Squeezing occurs for a variable at $45$ degrees where the wavefunction becomes  well defined in position. However, the effect of $r$ produces not such a good  well-defined localization of the wavefuction as the one for $r=0$.
It is clear that the Bohm and external potentials,  for both cases $r=0$ and $r\ne 0$, are given by time dependent coupled harmonic oscillators \cite{Guedes,Recamier,Iran}.\\
}

\section{Conclusions}
We have presented an operator method to engineering two-mode squeezed vacuum-like states; this method is based in the Madelung-Bohm approach to the Schrödinger equation and uses explicitly the Bohm potential. It is the correct selection of the phase of the wave function what allows us to find these entangled quantum states. The method provides an explicit form for the external time dependent quadratic potential that make possible the generation of the two-mode squeezed vacuum-like states.

\appendix
\section{Factorization of the evolution operator}\label{apa}
We propose as factorization of the evolution operator $\exp\left[\nu\left(\hat{a}^\dagger\hat{b}^\dagger-\hat{a}\hat{b} \right)  \right]$, given in \eqref{opsumII}, the ansatz
\begin{equation}
    \hat{U}(\nu)=e^{\nu(\hat{a}^{\dagger}\hat{b}^{\dagger}-\hat{a}\hat{b})}=e^{f_1(\nu)\hat{a}^{\dagger}\hat{b}^{\dagger}}
    e^{f_2(\nu)(\hat{a}\hat{a}^{\dagger}+\hat{b}^{\dagger}\hat{b})}e^{f_3(\nu)\hat{a}\hat{b}}.
\end{equation}
The derivative of the evolution operator with respect to $\nu$ gives, on the one hand,
\begin{equation}
    \frac{d\hat{U}(\nu)}{d\nu}=(\hat{a}^{\dagger}\hat{b}^{\dagger}-\hat{a}\hat{b})\hat{U}(\nu),
\end{equation}
and, on the other hand,
\begin{align}
    \frac{d\hat{U}(\nu)}{d\nu}&=\frac{df_1}{d\nu}\hat{a}^{\dagger}\hat{b}^{\dagger}e^{f_1(\nu)\hat{a}^{\dagger}\hat{b}^{\dagger}}    e^{f_2(\nu)(\hat{a}\hat{a}^{\dagger}+\hat{b}^{\dagger}\hat{b})}e^{f_3(\nu)\hat{a}\hat{b}}
    +\frac{df_2}{d\nu}e^{f_1(\nu)\hat{a}^{\dagger}\hat{b}^{\dagger}}
    (\hat{a}\hat{a}^{\dagger}+\hat{b}^{\dagger}\hat{b})e^{f_2(\nu)(\hat{a}\hat{a}^{\dagger}+\hat{b}^{\dagger}\hat{b})}e^{f_3(\nu)\hat{a}\hat{b}}
	\nonumber \\ &
    +\frac{df_3}{d\nu}e^{f_1(\nu)\hat{a}^{\dagger}\hat{b}^{\dagger}}
    e^{f_2(\nu)(\hat{a}\hat{a}^{\dagger}+\hat{b}^{\dagger}\hat{b})}\hat{a}\hat{b}e^{f_3(\nu)\hat{a}\hat{b}},
\end{align}
where, if  we use the identity $e^{t\hat{A}}\hat{B}e^{-t\hat{A}}=\hat{B}+t[\hat{A},\hat{B}]+\frac{t^2}{2!}[\hat{A},[\hat{A},\hat{B}]]+\frac{t^3}{3!}[\hat{A},[\hat{A},[\hat{A},\hat{B}]]]+\dots$, we obtain the set of coupled differential equations
\begin{equation} \label{eqs}
1=\frac{df_1}{d\nu}-2f_1\frac{df_2}{d\nu}+f_1^2\frac{df_3}{d\nu}e^{-2f_2}, \qquad
0=\frac{df_2}{d\nu}-f_1\frac{df_3}{d\nu}e^{-2f_2}, \qquad -1=\frac{df_3}{d\nu}e^{-2f_2},
\end{equation}
subject to the  initial conditions $f_1(0)=f_2(0)=f_3(0)=0$. It is not difficult to show that the solution of the above system of equations is
\begin{equation}
f_1(\nu)=\tanh \nu, \qquad f_2(\nu)=-\ln \cosh \nu, \qquad f_3(\nu)=-\tanh \nu.
\end{equation}


\begin{thebibliography}{99}
\bibitem{Schroedinger} Schr\"odinger, E. Die gegenw\"artige Situation in der Quantenmechanik. Naturwissenschaften 23, 807-812 (1935).
\bibitem{TMSV} Schumaker, B. L.; Caves, C. M.  New formalism for two-photon quantum optics. II. Mathematical foundation and compact notation. Physical Review A. 31, 3093–3111 (1985).
\bibitem{Prajapati} N. Prajapati, Z. Niu and I. Novikova, Quantum-Enhanced Two-Photon Spectroscopy Using Two-mode Squeezed Light, arXiv:2012.13745.
\bibitem{Yuen} H.P. Yuen, Phys. Rev. A 13 (1976) 2226.
\bibitem{Caves} C.M. Caves, Phys. Rev. D 23 (1981) 1693.
\bibitem{Loudon} R. Loudon and P.L. Knight, Squeezed light. J. of Mod. Opt. 34
(1987) 709.
\bibitem{TMarxiv} Fernando R. Cardoso, Daniel Z. Rossatto, Gabriel P. L. M. Fernandes, Gerard Higgins, Celso J. Villas-Boas, quant-ph, arXiv:2102.01032
\bibitem{Lewis} H.R. Lewis, Phys. Rey. Lett. 18 (1967) 510
\bibitem{Guasti} I. Ramos-Prieto, A. Espinosa-Zúñiga, M. Fernández-Guasti and H. M. Moya-Cessa. Quantum harmonic oscillator with time dependent mass
Modern Physics Letters B, 1850235 32, 1850235 (2018).
\bibitem{Pedrosa} I.A. Pedrosa and I Guedes,Int. J. 01 Mod. Ph.ysB. 18 (2004) 1379.
\bibitem{Ray} J. R. Ray, Phys. Rev. A 22 (1982) 729.
\bibitem{Guedes} 1D. X. Macedo and I. Guedes, J. Math. Phys. 53, 052101 (2012).
\bibitem{Recamier} H. M. Moya-Cessa and J. R\'ecamier. Comment on 'Time-dependent coupled harmonic oscillators.' [J. Math. Phys. 53, 052101 (2012)], Journal of Mathematical Physics 61, 114101 (2020).
\bibitem{Iran} I. Ramos-Prieto, J. R\'ecamier and H.M. Moya-Cessa. Time-dependent coupled harmonic oscillators: Classical and quantum solutions
International Journal of Modern Physics E 29, 2075001 (2020).
\bibitem{Asav} Warit Asavanant, Kan Takase, Kosuke Fukui, Mamoru Endo, Jun-ichi Yoshikawa, Akira Furusawa,  quant-ph, arXiv:2102.02359.
\bibitem{madel} E. Madelung, Z. Physik {\bf 40}, 322 (1927).
\bibitem{bohm} D. Bohm, Phys. Rev. {\bf 85}, 166 (1952).
\bibitem{durr} Dürr D. and Teufel S., \textit{Bohmian Mechanics. The Physics and Mathematics of Quantum Theory}. Springer-Verlag Berlin Heidelberg 2009.
\bibitem{Wheeler} W.P. Schleich and J.A. Wheeler, 
Oscillations in photon distribution of squeezed states, J. Opt. Soc. Am. A 10, 1715-1722 (1987)
\bibitem{Urzua} A.R. Urzúa, H.M. Moya-Cessa, I. Ramos-Prieto and  M.F. Guasti. Solution to the Time Dependent Coupled Harmonic Oscillators Hamiltonian with Arbitrary Interactions. Quantum Reports 1, 82-90 (2019)
\bibitem{Mehler} Mehler, F. G., "Ueber die Entwicklung einer Function von beliebig vielen Variabeln nach Laplaceschen Functionen h\"oherer Ordnung", Journal f\"ur die Reine und Angewandte Mathematik 66: 161–176, (1866).
\bibitem{erdelyi} Erdélyi, Arthur; Magnus, Wilhelm; Oberhettinger, Fritz; Tricomi, Francesco G. (1955), Higher transcendental functions. Vol. II, McGraw-Hill (scan:   p.194 10.13 (22))
\bibitem{stone} Mathematics for physics. A guided tour for graduate students. M. Stone and P. Goldbart 2009. 0521854032, page 65.
\bibitem{andrews} Larry C. Andrews, Ronald L. Phillips; Mathematical Techniques for Engineers and Scientists 0819445061, page 485. Copyright © 2003 The Society of Photo-Optical Instrumentation Engineers. SPIE Press monograph 2003.
\end{thebibliography}
\end{document}